# Energy Efficiency of Many-Soft-Core Processors

David Castells-Rufas
Universitat Autònoma de Barcelona
Edifici Enginyeria, Campus UAB
08193 Bellaterra, Spain
david.castells@uab.cat

Albert Saa-Garriga
Universitat Autònoma de Barcelona
Edifici Enginyeria, Campus UAB
08193 Bellaterra, Spain
albert.saa@uab.cat

Jordi Carrabina
Universitat Autònoma de Barcelona
Edifici Enginyeria, Campus UAB
08193 Bellaterra, Spain
jordi.carrabina@uab.cat

## ABSTRACT
The growing capacity of integration allows to instantiate hundreds of soft-core processors in a single FPGA to create a reconfigurable multiprocessing system. Lately, FPGAs have been proven to give a higher energy efficiency than alternative platforms like CPUs and GPGPUs for certain workloads and are increasingly used in data-centers. In this paper we investigate whether many-soft-core processors can achieve similar levels of energy efficiency while providing a general purpose environment, more easily programmed and allowing to run other applications without reconfiguring the device. With a simple application example we are able to create a reconfigurable multiprocessing system achieving an energy efficiency 58 times higher than a recent ultra-low-power processor and 124 times higher than a recent high performance GPGPU.

## 1. INTRODUCTION

Energy efficiency of computing systems is often measured by the amount of operations per second that the system can do within a power budget as described in the expression (1). We use G, for greenness to denote energy efficiency. Greenness is often expressed in operations per Watts, or one of its multiples. The Green500 list, for instance, uses GFLOPS/W, i.e. the number of $10^9$ floating point operations per second that can be executed with a Watt of electrical power. Since power times time equals energy, the expression can be rewritten as (2), meaning that G can be also understood as the number of operations that can be done with a Joule of energy.

$$G = \frac{Op}{T}\frac{1}{P} \qquad (1)$$

$$G = \frac{Op}{E} \qquad (2)$$

Computing systems have increased their energy efficiency significantly since their introduction. First electronic computers were implemented using vacuum tubes, having a large power consumption. In 1951, UNIVAC I was able to perform 2 kilo instructions per second (KIPS), while consuming 125 KW. It had no floating point support, which was added in following machines. But, in order to compare it with other machines, we assume an overhead of 100 integer instructions to execute a floating point instruction. Thus, our estimating is that its energy efficiency would be around $1.6 \cdot 10^{-12}$ GFLOPS/W

The introduction of the transistor allowed for much more energy efficient computers. In 1965, IBM System/360 91 was able to perform 1.9 MFLOPS within a 74 KW power budget, giving an energy efficiency of $25.7 \cdot 10^{-9}$ GFLOPS/W.

The change from bipolar transistors to CMOS transistors provided another significant gain, although it took some time to be fully adopted. In 1996, an Intel 80486DX2 was able to provide 2.6 MFLOPS while consuming 4 W, giving an energy efficiency of $0.6 \cdot 10^{-3}$ GFLOPs/W.

During the last 25 years the world top computers have increased their energy efficiency three orders of magnitude (see Figure 1). The number one system in the green500 list (as Nov. 2015) [1] has an energy efficiency around 7 GFLOPs/W.

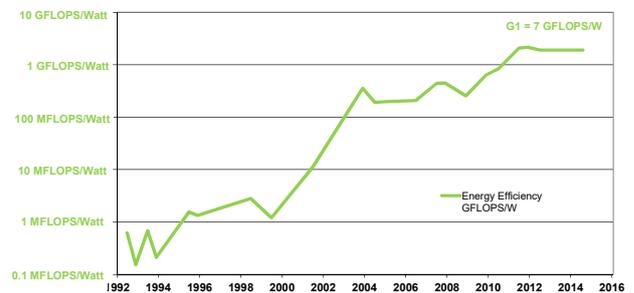

**Figure 1 Evolution of energy efficiency of top500 systems**

We are still far from the energy efficiency found in biological systems. Our brain, for instance, has an estimated equivalent processing capacity of 2.2 petaflops and consumes about 20W [2], giving an energy efficiency close to $1 \cdot 10^5$ GFLOPS/W. DNA computing could theoretically achieve an energy efficiency of $2 \cdot 10^{10}$ GFLOPS/W [3].

As the capacity of transistor integration is still growing but frequency gain has halted due to the high thermal density, industry has moved towards multi and many-core processors seeking to maintain and increase performance by exploiting thread-level parallelism and reducing frequency, and voltage if possible.

Many-core processor companies are marketing peak energy efficiency ratios above 10 GFLOPS/W. For instance, Adapteva reports up to 70 GFLOPS/W for its Epiphany IV many-core working at 800 MHz. In [4], authors actually claim to be able to get up to 9 GFLOPS/W.

Some FPGA vendors market a peak energy efficiency of 80 GFLOPS/W [5]. To implement systems that are able to reach high efficiency levels, FPGA vendors provide HDL and, recently, OpenCL toolchains to create application specific architectures to exploit the parallelism of the applications.

However, these approaches lead to specific solutions that cannot be used for other workloads unless the device is reconfigured. On the contrary, we are investigating the level of performance efficiency that can be reached by using a more general approach based on many-soft-cores, i.e. replicating soft-core processors and using parallel programming techniques.

## 2. ENERGY EFFICIENCY DRIVERS

The energy efficiency equation (1) takes operations per second over power consumption. Operation is an ambiguous term that can be used to denote whatever operation we are interested in. When used as a synonymous of a microprocessor instruction, a problem of indeterminism arise as different instructions usually take different amounts of time to complete. Different latencies for different instructions are inevitable but computer architects struggle to make different instructions throughput as similar as possible by using pipelining, branch prediction, cache memories, and other advanced techniques. In order to have fair comparisons between different systems relative results of benchmark applications are used instead of taking absolute numbers.

Whatever operations we measure, they can be expressed as a function of operations per cycle (OPC) and the clock frequency, as defined in (3)

$$\frac{Op}{T} = \text{OPC} \cdot f_{clk} \qquad (3)$$

The other element in (1) is power consumption. Power consumption in digital CMOS systems is approximately determined by the expression (4), which add the contributions from dynamic power (on the left) and the contribution from static power (on the right).

$$P = N \cdot \alpha \cdot f_{clk} \cdot C \cdot V^2 + N \cdot V \cdot I_{leakage} \qquad (4)$$

Combining (3) and (4) the (1) expression can be rewritten as (5), where $N$ is the number of transistors of the circuit, $\alpha$ the activity factor, $V$ the supply voltage, $C$ the effective load capacitance, and $I_{leakage}$ the leakage current of transistors.

$$G = \frac{OPC}{N \cdot \left( \alpha \cdot (C \cdot V^2) + \frac{V \cdot I_{leakage}}{f_{clk}} \right)} \qquad (5)$$

The only parameters with a positive correlation with the energy efficiency factor are $OPC$ and $f_{clk}$. Executing more operations per cycle, if power consumption and frequency are kept constant, will increase the efficiency of the system in a linear way. All other parameters appear in the denominator of (4); and all, except $f_{clk}$, have a negative correlation with the energy efficiency factor.

The parameter $\alpha$ indicates the probability of transitions occurring on transistors. The higher its value, the less energy efficient a system will be as it contributes to increase power consumption. Nevertheless, it is unlikely that the value of $\alpha$ goes to extremes, as it is typically between 10% and 20%, and the dark silicon effect [6] will help to make it even smaller. Similarly the number of transistors of a design also contribute negatively to energy efficiency. More transistors increase the total charge that causes dynamic power consumption, and also increase the silicon areas affected by leakage currents. FPGAs have a fixed number of resources, and all of them (including unused ones) would contribute to static power unless some measure is taken to mitigate this effect. Modern FPGAs used advanced techniques to reduce those undesired contributions [7].

Leakage current also affects negatively to energy efficiency as it increase the static power consumption. Again, some FPGA manufacturers [7] allow reducing it by modifying the back bias voltage of transistors.

Reducing supply voltage is a well-known technique to reduce power consumption and increase energy efficiency as it contributes quadratically in the denominator; however, reducing it is not straightforward because it cannot be lower than the transistor threshold voltage and, often, the frequency must also be reduced to permit a correct operation of the system. In addition, commercial FPGAs platforms usually work with a fixed supply voltage.

Clock frequency is positively correlated with energy efficiency. This could seem counter intuitive, since reducing frequency is often perceived as a power saving strategy working in conjunction with voltage gating or scaling (as in [8]). This is often the case when no more computing is needed, and there is no option to switch off the device or finish the task. This situation occurs, for instance, in a MPEG decoder that has already rendered a frame and is waiting for the appropriate time to display it in order to satisfy the specified frames per second (such as in [9]).

But generally, increasing clock frequency allows the task to finish faster. Thus, making the contribution of static power proportionally smaller. This observation was already made in [10] but is often misunderstood.

If we ignore static power, energy efficiency would be defined by equation (6). Static power has an increasing importance in new technology nodes, and should not be ignored, but measuring dynamic power is easier and is still the principal component when trying to reach peak performance [11].

$$G_{dyn} = \frac{OPC}{N \cdot \alpha \cdot C \cdot V^2} \qquad (6)$$

## 3. SOFT-COREs vs. HARD-COREs

The possible higher performance and better energy efficiency of FPGAs with respect to CPUs and GPGPUs has been already demonstrated by several works (see Table 1). Although FPGAs are not always the best option in terms of performance, they often become the best energy efficient alternative. A usual argument against FPGA implementations is the complexity in developing complete solutions using HDL languages. On contrast, high performance applications running on CPUs are easily programmed with high level programming languages like C/C++. GPGPUs use languages like CUDA and OpenCL that are based on C/C++ and introduce some annotations in the code to guide its execution to benefit from the coprocessing platforms.

**Table 1 Relative performance and energy efficiency of FPGAs/CPUs/GPGPUs**

|  | Relative Performance | | | Relative Energy Efficiency | | |
|---|---|---|---|---|---|---|
|  | FPGA | CPU | GPGPU | FPGA | CPU | GPGPU |
| [12] | 1 | 1/25 | 1/9 | 1 | 1/37 | 1/35 |
| [13] | 1/2.12 | 1/9.75 | 1 | 1 | 1/24.5 | 1/8.75 |
| [14] | 1 | 1/544 | 1/10.8 | 1 | 1/336 | 1/16 |
| [15] | 1/3.15 | 9.81 | 1/3.24 | 1 | 1/3.3 | 1/13.3 |

It is not straightforward for a programmer to exploit the peak performance of GPGPUs because programming style must be very aware of the particularities of the GPGPU architecture, like its SIMT execution style and memory hierarchy.

The skills needed for an optimal implementation in FPGA are even higher since designers have the challenge to create a complete new architecture from scratch using HDL and map the algorithm effectively on it. Latest efforts to provide OpenCL toolchains and High Level Synthesis (HLS) have eased the implementation of new designs on FPGAs, however the performance and efficiency of the designs implemented with this techniques are usually far from optimal.

## 3.1 Measuring Energy Efficiency of a Synthetic Benchmark

In this paper we want to study the energy efficiency that we can get with FPGA-based multiprocessing systems. We will start by analyzing the efficiency of a single soft-core processor.

If we use the logic resources of an FPGA to create a processor and execute a benchmark application on it, what energy efficiency can we expect with respect of the energy efficiency achieved by hard-core processors? The flexibility offered by the FPGA platform must have an impact, which, on the power consumption of basic logic, was measured to be around a factor 14 using the same technology node [16] for simple logic synthesis circuits.

A soft-core processor, though, is a more complex circuit which usually combine different types of resources, such as memory and DSP elements. Moreover, we will not compare soft-cores with their hardened implementation in ASIC, but with different energy efficient hard-core processors. In particular, we will compare a NIOSII implemented in a Stratix IV 530K device, with Cortex-A8 embedded in OMAP3530, and Intel i7-5500U.

In order to have a fair comparison, we should try to ensure that the conditions are similar for all platforms. In the NIOSII platform we are going to work with on-chip memory. This means that no external memory accesses will contribute to increase the power consumption. To have a similar environment on the hardened processors we use the Dhrystone benchmark [17]. It is so small that we can be sure that both code and data fit on the cache memories of the A8 and i7 processors and very few external memory accesses will be made.

The results of executing this benchmark on all platforms are shown in Table 2. The i7 processor has the best performance but a comparable energy efficiency with the A8 processor. The NIOSII soft-core appears to have a 5 times smaller energy efficiency and a 130 times smaller performance compared with the i7.

**Table 2 Results of Dhystone benchmark running on i7, A8 processors vs. a NIOSII instance on EP4SGX530**

|            | i7-5500U    | Cortex-A8   | NIOS-II     |
|------------|-------------|-------------|-------------|
| $Op/T$     | 20.028 GIPS | 0.883 GIPS  | 0.153 GIPS  |
| $P_{dyn}$  | 11.812 W    | 0.5 W       | 0.410 W     |
| $G_{dyn}$  | 1.69 GIPS/W | 1.76 GIPS/W | 0.37 GIPS/W |

We analyze the processor parameters to try to find a reason that explains why the energy efficiency of the NIOSII is lower but not as low as predicted in [16]. Table 3 shows some of the parameters that influence the performance and the energy efficiency of the systems obtained with expression (6).

Based on the reported transistor density, die size, and core dimensions we estimate that the number of transistors is around 14 MTr. for the A8 and 106 MTr. for the i7. The implemented NIOS-II system uses about 7 K Logic Cells (LCs) and $10^6$ memory bits. As a LC typically uses around 1000 transistors [20] and a memory bit uses 4 transistors, the design would be approximately equivalent to 43 MTr.

**Table 3 Some of the parameters that influence performance and power efficiency of the analyzed processors.**

|           | i7-5500U | Cortex-A8 | NIOS-II |
|-----------|----------|-----------|---------|
| $f_{clk}$ | 2.9 GHz  | 600 MHz   | 160 Mhz |
| $OPC$     | 6.91     | 1.47      | 0.96    |
| $V$       | 1.25 V   | 1.35 V    | 0.9 V   |
| $N$       | 106 MTr. | 14 MTr.   | 43 MTr. |

Unfortunately, it is impossible to accurately estimate the values for $\alpha$ and $C$. However, we can take its product (which we coin as $K_{tec}$) and infer it using the expression (7). This value can help us to elucubrate and understand the observed differences among processors.

$$K_{tec} = \frac{OPC}{G_{dyn} \cdot V^2 \cdot N} = \alpha \cdot C \quad (7)$$

Table 4 shows the $K_{tec}$ values obtained from the different processors, and their relative value compared to the value of A8. The i7 processor shows a smaller relative value that could be expected as being implemented in much more advanced technology node. On the other hand the NIOSII processor shows a worse relative value even being implemented on the next generation technology node.

**Table 4 Estimated effective capacity and switching activity product by processor**

|                  | i7-5500U          | Cortex-A8         | NIOS-II           |
|------------------|-------------------|-------------------|-------------------|
| Process node     | 14nm              | 65nm              | 40nm              |
| $K_{tec}$        | $2.46 \cdot 10^{-17}$ | $3.26 \cdot 10^{-17}$ | $7.36 \cdot 10^{-17}$ |
| Rel. $K_{tec}$   | 0.75              | 1                 | 2.25              |

To shed some more light about the processor differences, we can try to remove the capacitance reduction caused by the evolution of process node technologies. As estimated in [18], the 40nm process node should contribute to reduce $C$ by a factor of 0.68 with respect to the 65nm process node, while the 14nm process node should contribute by a factor 0.47. So by removing this effects we get the values shown in Table 5.

The reduction $C$ provided by new advanced technology nodes might be dilapidated by the high frequency target used in the synthesis process of the i7, which will inevitably increase the value of $C$ as explained in [19]. The high transistor count of the i7 is the price paid to create a superscalar architecture that is able to obtain a high OPC value. Comparing it with the A8, the transistor count and bigger synthesis effort is compensated by the high OPC and the smaller capacitance of the new technology node, finally getting a very similar $G_{dyn}$ factor.

**Table 5 Process node contribution to capacitance and its relative value with respect to A8**

|  | i7-5500U | Cortex-A8 | NIOS-II |
|---|---|---|---|
| $C_{pn}$ (from [18]) | 0.47 | 1 | 0.68 |
| Rel. $K_{tec}/C_{pn}$ | 1.6 | 1 | 3.3 |

On the other hand the energy efficiency of the NIOS II processor is almost five time worse than that of the other processors. This is the result of having a transistor count overhead to implement the reconfigurable fabric, but also a higher capacitance due to the reconfigurable interconnect.

## 3.2 Soft-Core Performance

Performance in soft-cores is also limited as Soft-Cores suffer from a reduced frequency of operation. Although [16] reports only a 3 or 4 penalty factor for FPGA speed compared to ASIC, this relatively small factor is often only achieved for very small designs. The typical factor for complex designs is often higher, in the order of 18 to 30 such as in [21] and [22].

Although soft-cores do not shine for their performance or energy efficiency they have something unique to offer: flexibility. They can be modified and complemented in order to achieve higher OPC trying to consume few extra transistors. Increasing OPC shall offer both higher performance and energy efficiency.

In addition, since modern FPGAs have an enormous number of resources, a large number of soft-cores can be implemented to exploit thread-level parallelism of applications (such as in [23]) and try get to higher levels of performance.

## 4. ILLUSTRATING EXAMPLE

In order to test how soft-cores can be modified to achieve higher levels of energy efficiency we implement a very simple application that computes the number of prime numbers in the first $10^6$ numbers avoiding the Sieve of Eratosthenes algorithm.

This is an embarrassingly parallel application that can be used demonstrate how both instruction and thread level parallelism can be exploited in many-soft-cores. The algorithm is based on a brute force approach to determine which numbers are primes. Basically, all numbers in the range must be evaluated for being prime. The function to determine if a number is prime can be described with the following code:

```
int isPrime(int v)
{
   int i;

   if (v >= 0 && v <= 3) return 1;

   if (v % 2 == 0) return 0;

   for (i = 3; (i*i)  < v; i++)
   {
      if (v % i == 0)
         return 0;
   }
   return 1;
}
```

Indeed, the application has a lot of parallelism. All the loops of the application can be parallelized since they do not have any data dependency. A possible simple parallelization using OpenMP is shown in the following code.

```
#pragma omp parallel for private(i,n) shared(primes)
for (i = 2; (i*i) < n; i++)
{
   if (isPrime(i))
   {
      #pragma omp atomic
      primes++;
   }
}
```

After compilation in gcc with full optimization (-O3), the single thread application takes 51 seconds on the i7-5500U processor. The OpenMP version reaches a maximum speedup factor of 2.3 on the same processor.

### 4.1 Soft-Core Baseline Implementation

We will follow an iterative method starting with the very same application that we will refine in order to try to reach higher levels of energy efficiency.

The baseline system consist on a simple NIOS II system running on 60 MHz and implement the single thread application using the gcc tool-chain and full optimization. The system is synthesized for the EP4SGX530 FPGA. As expected, the performance of the system is very poor, 630 times worse than the laptop version, but the energy efficiency is just an order of magnitude worse.

### 4.2 Adding an Application-Specific Custom Instruction

Analyzing the source code of the application we can detect opportunities to create custom hardware that could allow us to increase the instruction parallelism, and improve performance and energy efficiency. Figure 2 shows the data flow graph of the instructions executed in the loop of the *isPrime* function, which is the obvious hot-spot of the application.

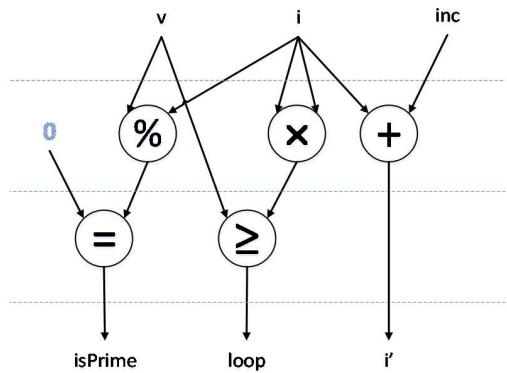

**Figure 2 Data-Flow graph of the operations executed in the loop**

To try to improve the performance of the *isPrime* function we implement the loop body on a custom instruction that receives the number to check whether it is prime as an argument, and returns a *boolean* value informing the result.

We implement the logic described in Figure 2 as a logic block in JHDL [24] for faster validation, export it to Verilog for later synthesis. We call the block *Iter*. Similarly we implement a NIOS

Custom Instruction with the design shown in Figure 3. The operation is very simple, the value to check is passed as the *a* argument to the custom instruction and it is registered. An *i* register is initialized with value 2 and is passed to the *Iter* module, which keeps incrementing it at every iteration. If *v* is divisible by *i* the *isPrime* signal is asserted. If not the loop signal indicates whether a next iteration should be performed with the next value of *i*. A simple FSM controls the data path.

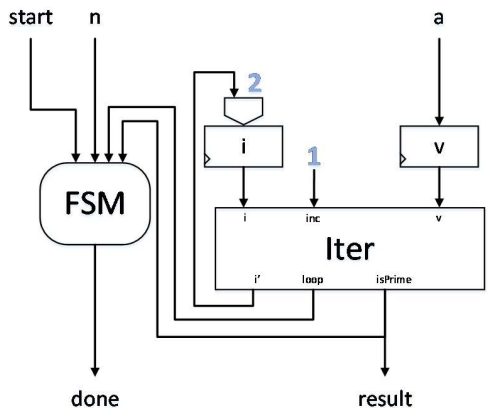

**Figure 3 Custom Instruction design**

The design is synthesized and integrated into the former NIOS II system. The application must be modified to take profit of the new custom instruction. The new system shows a speedup of x7 approx. compared with the previous one, while the energy consumption is almost the same. This increases the energy efficiency to a similar to that of the i7 processor. Although the performance is much lower, 274 times worse.

## 4.3 Multiunit

The loop of prime number computing can be unrolled nicely since there is no data dependencies between iterations. Unrolling means to concurrently execute some of the different iterations of the loop. This could be done by replicating the *iter* loop units and assigning different values of *i* to each one.

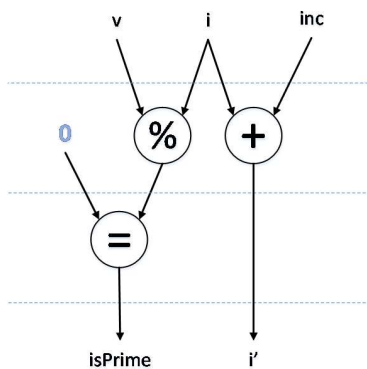

**Figure 4 Data-Flow graph of the optimized loop unit**

But the nature of the algorithm allows further optimization. Is not necessary to do the check for the loop bounds in every iteration, by doing it in blocks the behavior is not altered and some hardware resources can be saved. To implement this idea, a new optimized unit without the hardware resources to check the bounds is done (see Figure 4).

Figure 5 depicts the new custom instruction, which combines the former iter unit with the new optimized one to save some resources. A system with 10 units is build following this strategy. Like in previous design, the system is easily implemented and verified in JHDL.

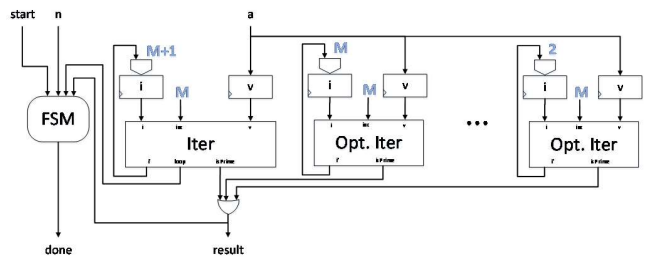

**Figure 5 New multiunit custom instruction**

The new design is synthesized an included in the NIOSII system. The application must be modified again to use the new custom instruction. When executing the benchmark we see that power consumption is slightly increased due to the additional resources taking part on the design. However, performance is increased by almost a factor x10. This boosts the energy efficiency to a higher level than that of the laptop processor.

## 4.4 Exploiting Pipelining

The system is still underutilized, since the divider units have a latency of 5 cycles due to its pipelined design, but just a single stage of the pipeline is active per cycle. In addition, some additional cycles are used by the control unit. It seems possible to increase the utilization of the dividers pipeline.

Again, the algorithmic nature of the problem allows it. The value of *i* could be increased every cycle without the need to wait for the completion of the previous modulo operation. All the iterations are done to find if any number give a zero remainder. So, it is not important whether the detection takes some additional latency or the reference to the number that produced the result is lost. Thus, a new value of *i* can be injected in each unit every cycle. The latency of the detection is increased but it can be easily handled by the control unit.

After implementing this change the system has to be synthesized again and the application has to be modified and recompiled to use the new custom instruction. Again, the increase in power consumption is minimal since we are just adding a small amount of logic to use the resources more efficiently. The speedup factor vs. the previous version is almost x6, which is in line with the number of cycles that were wasted by underutilizing the dividers pipelines. The performance efficiency factor is boosted again by a similar factor.

The system has a comparable performance to the i7 processor, but its energy efficiency is 23 times better that the best implementation with 8 threads on the multicore standard processor.

## 4.5 Parallel Architectures

To continue increasing the performance of the system several strategies could be followed. More parallel units could be added to the custom instructions, and there would be also the possibility to increase the pipeline levels of the divider. The presented designs use dividers with a pipeline of 5 stages, which certainly limits the maximum frequency of the system. It is reasonable to increase the pipeline levels up to 35 to reach a higher frequency of operation. Having this long latency would increase the opportunity to both increase the frequency and still benefit from the new levels of pipeline.

Another option to increase performance is to replicate the processor with its custom instructions and implement a parallel version of the application. The initial computer application was parallelized using OpenMP. We still have no support for OpenMP on the reconfigurable multiprocessors that we build, but we support threads [18]. So, we created an 8-core multi-soft-core that includes the best performing custom instruction and use a multithreaded version of the benchmark application.

The resource usage of the created system is detailed in Table 6. Notice that only a 26% of the LUTs of the device are used, so it would be perfectly possible to create a system with 24 cores and expect a linear speedup of the system.

As easily seen in Figure 6, the resource usage is dominated by the custom instruction units we have created (shown in blue colors).

**Table 6 Synthesis results of many-soft-core system on EP4SGX530**

| Element | LUTs | FFs | Memory bits | DSP |
|---|---|---|---|---|
| CPU 0 | 1419 | 1600 | 63104 | 4 |
| CPU 1-7 | 1200 | 1305 | 20224 | 4 |
| FPU 0-7 | 1343 | 1254 | 4608 | 20 |
| CI 0-7 | 10474 | 1817 | 0 | 0 |
| Other | 6811 | 6773 | 9528320 | 0 |
| TOTAL | 111190 (26%) | 42401 (10%) | 9769856 (46%) | 192 (19%) |

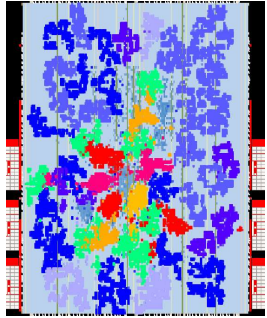

**Figure 6 FPGA Floorplan of the 8-core multiprocessor on the EP4SGX530**

After synthesis, compilation and execution on the platform, the execution time is reduced to 3.97 seconds when running with 8 threads, while achieving an energy efficiency factor 58 times higher than the i7.

## 5. RESULTS REVIEW

We have created a FPGA multiprocessor system based on soft-core processors that have been slightly modified to give a high performance and high energy efficiency platform. Before measuring the number of operations per second of the benchmark on various systems, we obtained the aggregate value of all the operations performed in each iteration of the loop inside *isPrime*. Thus, this number representative of the total number of operations of the algorithm.

The best energy efficiency obtained in the i75500U was 27.7 MOPS/Watt using 6 threads (see Table 7). The energy efficiency increased with the number of threads but not more than a factor 1.5.

**Table 7 Performance and Energy Efficiency results of the OpenMP version executing in i7-5500U**

| Threads | Time (seconds) | $P_{dyn}$(Watts) | Estimated MOPS/Watt |
|---|---|---|---|
| 1 | 51.42 | 10.7 | 17.571 |
| 2 | 39.77 | 15.6 | 15.582 |
| 3 | 31.33 | 14.6 | 21.135 |
| 4 | 27.45 | 15.3 | 23.018 |
| 5 | 25.63 | 15.6 | 24.179 |
| 6 | 23.89 | 14.6 | 27.716 |
| 7 | 22.92 | 16.6 | 25.409 |
| 8 | 22.35 | 16.6 | 26.057 |

The same application was implemented in NIOSII. The first baseline version had a terrible performance due to the low frequency of operation of the system. However, as we created more custom logic the execution time was reduced below the execution time of the i7 processor while achieving an energy efficiency orders of magnitude higher than the laptop processor (see Table 8).

**Table 8 Single Processor NIOS II designs**

| Design | Time (seconds) | $P_{dyn}$ (Watts) | Estimated MOPS/Watt |
|---|---|---|---|
| Baseline | 14093 | 0.28 | 2.45 |
| Custom Instruction | 1879 | 0.3 | 17.146 |
| 10 units | 188.19 | 0.4 | 128.42 |
| Pipelining | 31.63 | 0.5 | 611.27 |

Then, a multiprocessor system was created with 8 cores, supporting threads programming model. When running with a single thread the performance efficiency was reduced because the additional logic was contributing to a higher power consumption. But very soon the additional logic contribution to power consumption was almost irrelevant in comparison with the performance and efficiency gains.

**Table 9 Multiprocessor NIOSII design**

| Threads | Time (seconds) | $P_{dyn}$(Watts) | Estimated MOPS/Watt |
|---|---|---|---|
| 1 | 31.63 | 0.75 | 407.51 |
| 2 | 15.55 | 0.9 | 690.76 |
| 3 | 10.32 | 1 | 936.75 |
| 4 | 7.57 | 1.1 | 1160.95 |
| 5 | 6.35 | 1.2 | 1268.67 |
| 6 | 5.27 | 1.35 | 1358.81 |
| 7 | 4.53 | 1.4 | 1524.33 |
| 8 | 3.97 | 1.5 | 1623.39 |

Finally, we implemented a CUDA version of the benchmark and executed on the GK110-400. It uses 28 blocks with 1024 threads each, to a total of 28672 threads. The performance on GPGPU was just a little bit below that of the multi-soft-core system, but its energy efficiency was even worse than that of the i7 processor.

**Table 10 GPGPU NVIDIA GK110-400**

| Time (seconds) | $P_{dyn}$(Watts) | Estimated MOPS/Watt |
|---|---|---|
| 3.91 | 189 | 13.082 |

In summary, the best performance achieved by the multi-soft-core processor is 5.6 times better than that of the computer, but the energy efficiency is 58 times larger. Moreover the scalability profile for the application is almost ideal (see Figure 7), and the use FPGA is far from being used at its maximum.

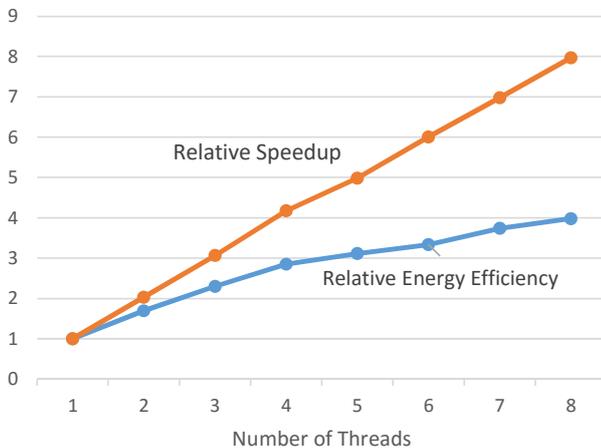

**Figure 7 Relative speedup and energy efficiency of multi-soft-core**

## 6. CONCLUSIONS

The number of processors that can be implemented in a FPGA is only limited by its resources. We have proven how augmented soft-core processors can be replicated and combined to form multi-soft-core and many-soft-core processors to reach high levels of energy efficiency and performance.

Using soft-core processors instead of OpenCL or HLS can let designers to invest a limited effort to create extra custom hardware that allows to increase the instruction level parallelism that is the key driver for energy-efficiency, and reuse all the parallel programming know-how to exploit the thread-level parallelism to increase the performance.

Using this strategy we have started with a single soft-core system having an energy efficiency of 2.45 MOPS/Watt and ended with a 8-core multiprocessor with 1623.39 MOPS/Watt. The former system achieve a 58 times better energy efficiency than a commercial ultra-low-voltage processor such as the i7-5500U. Compared with GPGPUs the multi-soft-core system achieves a similar performance but with a significant higher energy efficiency (124 times higher).

## 7. ACKNOWLEDGMENTS
This work was partly supported by the European cooperative CATRENE project CA112 HARP, the Spanish Ministerio de Economía y Competitividad projects IPT-20120847-430000 and TEC2014-59679-C2-2-R.